Stresa, Italy, 25-27 April 2007# HIGH EFFICIENCY 3-PHASE CMOS RECTIFIER WITH STEP UP AND REGULATED OUTPUT VOLTAGE - DESIGN AND SYSTEM ISSUES FOR MICRO GENERATION APPLICATIONS

*J.-C. Crebier, Y. Lembeye, H. Raisigel, O. Deleage, J. Delamare, and O. Cugat*
Laboratoire de Génie Electrique de Grenoble / G2Elab / www.g2elab.inpg.fr
Grenoble, FRANCE
crebier@g2elab.inpg.fr, lembeye@g2elab.inpg.fr
**ABSTRACT:**

This paper presents several design issues related to the monolithic integration of a 3-phase AC to DC low voltage, low power rectifier for 3-phase micro source electrical conditioning. Reduced input voltage operation (down to 1V), high efficiency, and output voltage regulations are implemented, based on commercially available CMOS technology. Global design and system issues are detailed. The management of start-up sequences under self supplied conditions as well as output voltage regulations are specifically addressed. Simulation results, practical implementation and validation are presented. They are based on the association of three micro elements: a 3-phase micro-generator, a stand alone 3-phase AC to DC integrated rectifier, and an output voltage conditioner based on a commercially available IC.
## 1. INTRODUCTION

In recent years, new micro-sources of electrical energy for wireless sensors and µ-devices have been investigated [1]. These new micro-sources, which represent an alternative to chemical batteries, generally provide a low-power electrical output, often low-voltage, and sometimes erratic. The associated power electronics providing the delicate interface between the load (sensor, transmitter) and its autonomous supply must be adapted to these specific requirements, the same time, it must be able to regulate and stabilize the DC voltage applied to the load. Additional requirements in design and system integration are minimal wastage of energy and limitation of device size [2].

The paper reports on the selection, design, fabrication and testing of a highly integrated, low power conditioner for 3-phase micro-generators [3]. Firstly, the choice of converter topology is investigated, having in mind that the converter must rectify the 3-phase variable input signals and at the same time, it must provide a regulated and stabilized boosted DC output voltage to the load. After this first section, we focus on the design of a stand-alone full-wave rectifier for 3-phase AC inputs in the range of 1-3.3 V (line-to-line), 5-500 mA at 10-100 kHz. Specific low powering issues such as low voltage rectification, self-powering and automatic start-up are clearly presented. These issues have been solved by original designs and concepts realized in CMOS 0.35 µm technology. Measurements on the fabricated µ-rectifier coupled to a 3-phase micro-generator indicates its full functionality and high conversion efficiency (up to 90 %). A final part presents the second stage of the conditioner, designed to boost and regulate the output voltage of the converter. Design issues are related, based on the use of a selection of commercially available devices. The global converter operation is validated and characterized through experimental implementations.

## 2. TOPOLOGY SELECTION FOR 3-PHASE RECTIFICATION AND OUTPUT VOLTAGE REGULATION.

Three-phase voltage rectification is a common, well known and well covered power electronics design topic. Usually based on rectifier diodes, for low-voltage low-power applications, these devices are being replaced by MOSFET transistors which are preferred for their lower forward voltage drops and their absence of recovery. Operating either in synchronous mode or coupled with intelligent drivers, threshold-less as well as reduced forward voltage drops are achieved. These full-wave converters perform AC to DC rectification without the capability to boost nor regulate the output voltage level. The basic topology is presented on figure 1, implemented with complementary MOSFET transistors driven by comparators.

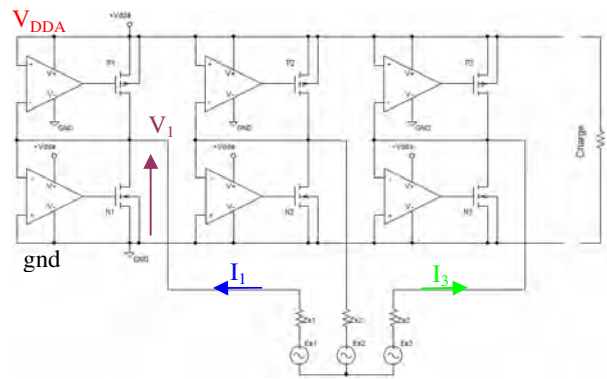

*Figure 1. Topology of a MOSFET-based full-wave 3-phase rectifier*

**2.1. Two stage cascaded topology.**

Output voltage boost and regulation may be carried out using a second stage power electronic or switched capacitor converter. Its choice depends in general on power flow level and range of conversion. Figures 2 presents two basic topologies used to perform voltage boost and regulation.

©EDA Publishing/DTIP 2007                                                                                          ISBN: 978-2-35500-000-3

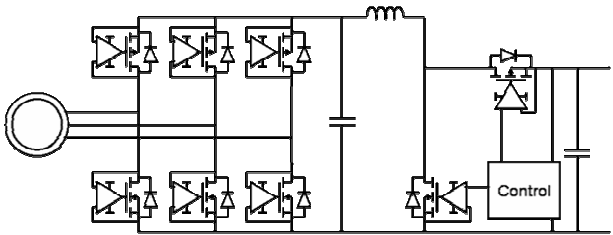

*Figure 2. Topology of dual- stage AC-DC and DC-DC step-up boot converter*

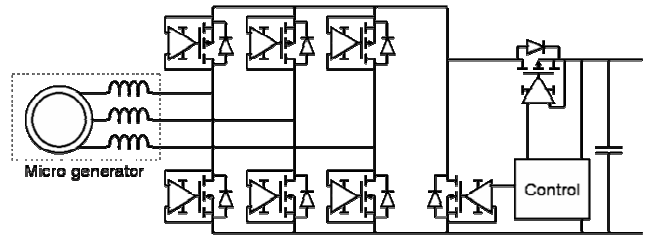

*Figure 3. Quasi-single stage power electronic topology for AC-DC step-up electrical conditioning.*

Nowadays, the active parts of this second stage can easily be integrated into a single die, and devices are commercially available [4, 5]. However, they also require several passive elements for storage and filtering which up to now remain difficult to integrate in or above IC [6, 7]. Luckily, thanks to these dual-stage cascaded topologies, the two converters are decoupled, which offers the opportunity to optimize each one separately. As a result, the second stage can operate at much higher switching frequencies, allowing drastic reductions in passive element sizes and stored energy levels ; indeed, the middle stage storage capacitor can be lowered thank to the second stage high bandwidth regulation loop.

**2.2. Single-stage topologies**

In order to reduce the number of converters but also the number of active and passive components, as well as to increase the global efficiency of the electrical conditioner, it may be interesting to merge the cascaded topology into a single-stage or quasi-single stage converter. In power electronics, extensive research activities have been carried out in these fields in the past decade. In order to perform AC to DC rectification as well as voltage boost and regulation, the single-stage converter must switch the input current at higher frequencies than the micro-generator's output AC signals. Taking advantage of its phase inductances which become dominant at high frequencies, energy can be stored and then released in a boost or step-up manner in order to amplify and to regulate the output voltage level. According to our application, two possible topologies have been identified.

The first topology is only based on the removal of the middle stage storage capacitor. Operating the boost chopper at high switching frequencies, and taking into account micro-generator phase impedances, the output voltage can be amplified, and at the same time the generator current waveforms can be improved (leading to reduced harmonic contents and reduced harmonic losses). Figure 3 presents the topology. When operating the boost part of the converter in discontinuous mode, the 3-phase AC-DC converter acts only as a switch-gear for the 3-phase current. All components can be integrated in the same silicon chip in CMOS technology, which is a highly suitable solution for global monolithic integration. Besides, fewer passive elements are needed. The main drawback is that the stress applied to the boost chopper is quite important. Besides, the converter voltage ratings must be set according to output voltage levels.

The second solution takes advantage of the technology used for the implementation of the 3-phase AC-DC rectifier based on threshold-less, reduced forward voltage drop MOSFET cells. Thanks to these components, active 3-phase rectification and step-up conditioning can be achieved, switching all transistors at much higher frequencies than the input signals generated by the micro-generator. The resulting single-stage topology is depicted on figure 4. In this topology, the number and sizes of active devices is minimized as well as the number of passive ones. In addition, the stress on active devices is equally shared among all transistors, making this solution highly suitable for our aims.

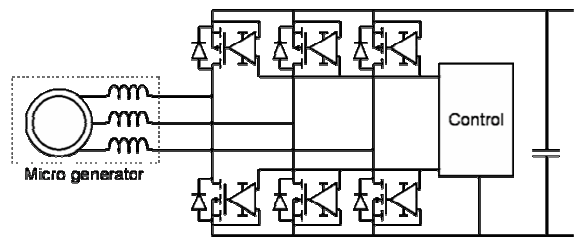

*Figure 4. Single-stage AC-DC boost rectifier topology.*

**2.3. Micro conditioner selection.**

The choice of the best converter relies on many design and operating issues. At first, the single-stage topology seems to offer the best compromises. However, it requires operating power CMOS inverter legs at high switching frequencies. This is quite challenging because the switching transition of the CMOS legs at high frequencies will generate large power losses making its optimization difficult.

Large CMOS components are not suitable for high switching frequencies with basic control, because the occurrence of short-circuits during transitions creates very large and undesirable source currents. As a result, in order to offer good results this topology requires extensive and specific research activity on CMOS switching transition management [8]. This is out of the scope of this paper and first work.

A quasi-single stage structure also seems interesting, but its implementation is delicate for the same reasons as above. Indeed, the boost chopper must operate in synchronous mode at high switching frequencies while carrying the full load current. In addition, all devices must be rated to handle higher voltage levels.

Although the cascaded solution based on a dual-stage topology requires a larger amount of active and passive devices, it appears as being the most affordable solution at the moment. Besides, the primary and secondary stages are decoupled, which allows for their respective optimizations. Finally, at the moment, a cascaded topology allows to maintain switching frequencies of all converters at reasonable levels, leading to a good global efficiency (the price to pay being the size and the amount of passives).

The following part of the paper will now focus on the global design and system integration issues related to the monolithic and hybrid integrations of the cascaded topology.

## 3. THREE-PHASE HIGH EFFICIENCY RECTIFIER WITH STEP-UP OUTPUT

This part presents the design of a dual-stage 3-phase AC-DC conditioner. According to the previous section, this function has been separated in two parts. The first part deals with the 3-phase rectifier that has been the subject of particular studies for its integration in CMOS technology, which are presented in the following paragraph. Then the fabrication of the second stage using a commercial circuit is presented.

### 3.1. Converter primary stage: the three-phase rectifier

The monolithic integration of the 3-phase AC to DC micro conditioner relies on several design and technological issues.

First of all, in order to minimize converter losses, a low voltage CMOS technology has been considered. It offers the possibility to implement "power" devices with adapted characteristics (low ON state voltage drop, low voltage thresholds) while giving the opportunity to design highly dedicated control electronics. The rectification is no longer implemented with diodes (not even Schottky): instead, we use MOSFETs designed and controlled to rectify the low level voltages without threshold and under reduced forward voltage drops. Specific care is given to the design of the power switches and their control. Each power switch is controlled by an associated comparator which drives it according to its voltage polarity (Figure 1). These comparators are supplied by the variable, low-level rectified output DC voltage (for general and stand-alone purposes). As a result, the rectifier is self-supplied and it must be designed to meet many specific functional requirements. In this paper we will focus on two specific functionalities that are required for this rectifier. First, the start-up sequence is presented and its influence on the design is discussed. Second, the difficult case of starting this self-powered circuit when having a heavy load (understand a large discharged capacitor tank) is treated and a solution is presented. The other functionalities of this 3-phase rectifier have already been presented in [9].

#### 3.1.1. Start-up sequence

Since the comparators are directly supplied by the rectifier output DC voltage, there is obviously a problem at start-up. Since no power is available at the converter output, the comparators are not supplied and the power switches are not driven. Thus, another specific circuit is needed in order to ensure start-up until the output voltage reaches 1V, during at least a few µs so as to stabilize the current and voltage references within the comparators. A first solution would be the use of the inherent body diodes within the MOS switches to rectify the voltages. In this case, all power MOSs must be maintained in their OFF state by using, for example, pull-down resistors connected to their gates. But in this case, due to diode voltage drop, starting the µ-rectifier requires an input voltage level higher than 2 x 0.6 V + 1 V = 2.2 V (voltage drop of two diodes in series + minimum output voltage to insure comparator minimum supply level) to enable the self-supplied operation. Moreover, the pull-down resistors would produce additional losses during the active drive of the µ-rectifier.

This problem has been solved by the addition of an autonomous electronic start-up circuit presented figure 5. During the output voltage shortage, this circuit controls directly the power switches instead of the comparators. The strategy of this control is based on the synchronous rectification technique. In fact, the input AC voltages are directly used to drive the MOS power switches (through transistors M3 and M4 on figure 5). Once the output rectified voltage reaches 1 V after a few µs, the active control by comparators is initiated and the synchronous start-up circuit is then automatically disconnected so as not to interfere with the active drive. Transistors M3 and M4 must be switched on at the beginning of the start-up sequence when:

- the phase-to-phase voltage $U_{12} > V_{TH\_P1}$ (0.7 V) > $V_{TH\_N2}$ (0.5 V),

- the rectified voltage ($V_{DDA}$-GND) ≈ 0 V; all capacitors are discharged,

- the reference current is not established $I_{REF} = 0$ A.

And, they have to be switched OFF at the end of the start-up sequence when the supply voltage and the reference current are established; VDDA ≥ 1 V, $I_{REF}$ ≈ 2 µA.

With these conditions, the circuit presented on figure 6 can be defined.

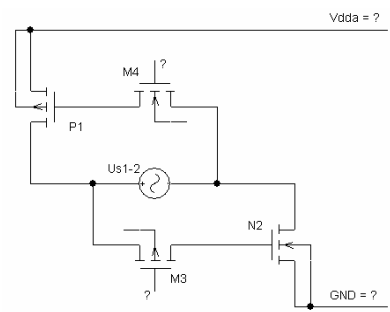

*Figure 5. Micro-rectifier start-up circuit topology. Transistors M3 and M4 are used to drive P1 and N2 as a synchronous rectifier.*

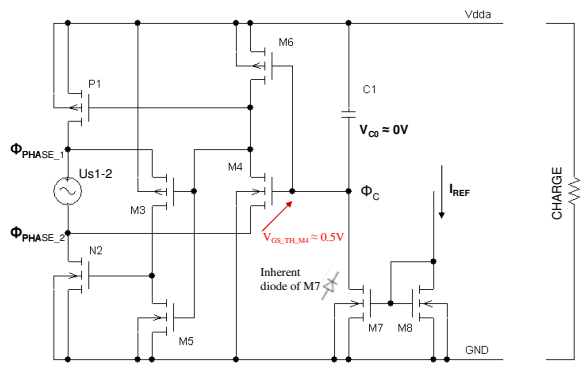

*Figure 6. Simplified diagram of the start-up circuit*

Note that in this circuit, all transistors work in switching mode (except M7 and M8). At first and as long as $U_{12} \geq 1V$, transistors M4, M3, P1 and N2 are switched ON if:

$$\Phi_C - \Phi_{PHASE\_2} > V_{TH\_M4} \approx 0.5V$$

Where $\Phi_C$ is the electric potential on the floating terminal of capacitor C1. Note that this potential is always between $\Phi_{PHASE\_MAX} > \Phi_C > \Phi_{PHASE\_MIN}$ because of very small but permanent current leakage.

Since P1 and N2 are ON, $V_{DDA}$ starts to rise up to:

$$V_{DDA} = U_{12} - V_{DS\_ON\_P1} - V_{DS\_ON\_N2}$$

Nevertheless, $V_{C1}$ remains unchanged and transistor N7 stays OFF until reference current -and thus the µ-rectifier commands- are established.

The start-up circuit maintains the output terminals of the µ-rectifier connected to $U_{12}$. Once the current reference is established, C1 starts to charge through M7 until $V_{C1}=V_{DDA}$ and $\Phi_C$ = GND. When $\Phi_C$ reaches GND, M4 and M3 are definitively switched OFF, thus ending the starting sequence so as to not interfere with the active drive.

The automatic start-up process and circuits are detailed in [10]. Note that, in steady-state operation, the active control performed by comparators is much more efficient than the synchronous one especially in 3-phase structures with sinusoidal inputs.

*3.1.2. Starting with heavy load.*

Nevertheless, when the µ-rectifier is loaded, the start-up circuit may not be sufficient since only a small part of the µ-generator power is converted at the output of the rectifier (only the $U_{12}$ positive half-wave is rectified). In this case, if the load connected below the rectifier is too large or if the filter capacitance is too heavy, this limited power may not be enough to establish a sufficient supply voltage for the active control command (Figure 7).

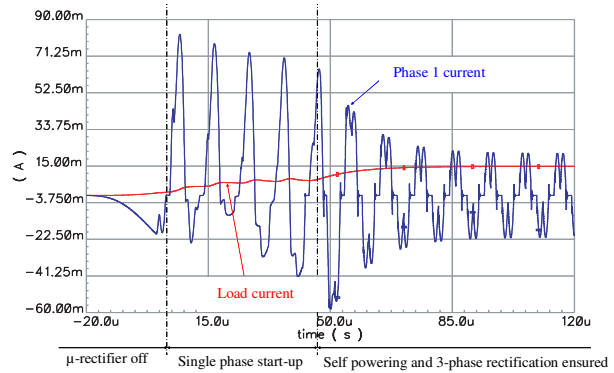

*Figure 7. Start-up simulation for an output capacitor of 1 µF.*

To overcome this difficulty, a "gndc" terminal has been added at the output of the µ-rectifier, and the output capacitor has been connected between this terminal and $V_{DDA}$ (Figure 8). During start-up, before the 3-phase active control command is activated, the "gndc" terminal is set to high-impedance in order to disconnect the output capacitor from the µ-rectifier.

As soon as the 3-phase µ-rectifier works correctly, the gndc terminal is progressively connected to the GND terminal allowing the smooth charge of the output capacitor making use of the full source power (Figure 9).

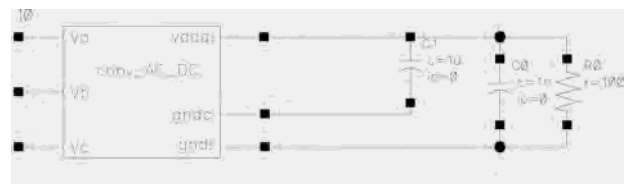

*Figure 8. µ-rectifier bloc diagram. After the start-up sequence, the gndc terminal is progressively connected to gnd!.*

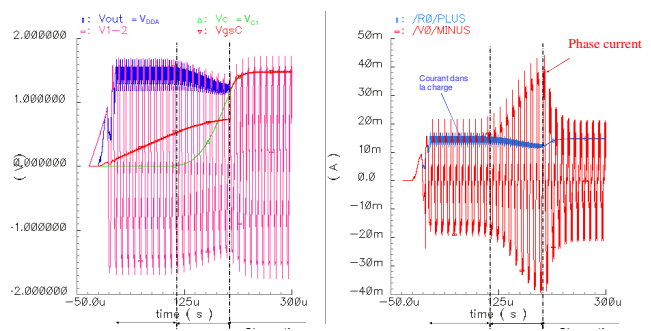

*Figure 9. Start-up simulation for an output capacitor of 1 µF progressively connected via the gndc terminal.*

The µ-rectifier start-up has been achieved using the simultaneous action of these two circuits. For light loads connected to the µ-rectifier, the start-up sequence lasts one half of the positive half-wave of $U_{12}$ phase-to-phase voltage applied to the input. Moreover, simulations carried out with Cadence have shown that our µ-rectifier can start with input voltage $U_{12MAX}$ as low as 1 V.

### 3.2. Converter secondary stage: the step-up.

The secondary stage must increase the µ-rectifier output voltage (1-3.3 V) up to 5 V. Therefore, it is necessary to dispose of a structure allowing: a) an important input voltage excursion swing and b) an important and adjustable output-to-input voltage ratio while having a high efficiency. These two characteristics do not match with those of charge pumps because charge pumps a) are generally designed for one specific voltage ratio, and b) offer high efficiency only if this ratio is observed. To perform the step-up stage, a boost converter was therefore chosen, and has been realised using the self-powered commercial circuit MAX1676. This component integrates the two active components (power MOS and power diode) built in MOS technology. According to the datasheet, this circuit can operate from 0.8 V input voltage, and output voltage can be adjusted from 2 to 5.5 V (5 V in our application). Static consumption of the circuit is as low as 16 µA and efficiency can reach 94% for a 200 mA output current. Only two additional components are needed: a power inductor and an output capacitor. The switching frequency is 500 kHz.

The inductor and capacitor were determined according to the datasheet, and figure 10 shows the diagram of the realised boost converter (L = 22 µH, C = 47 µF).

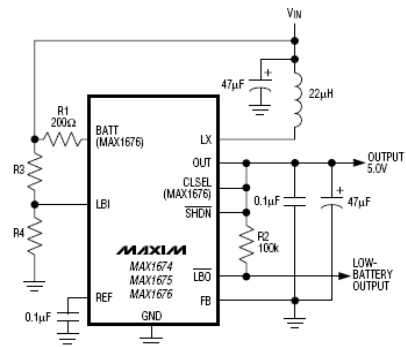

*Figure 10. Boost converter based on MAX1676.*

### 4. TEST OF AC-DC INTEGRATED µ-CONVERTER.

The µ-rectifier chip was designed and simulated with Cadence using the AMS design kit [11] (Figure 11). The fabrication was processed by the AMS foundry via the CMP society [12]. The chip (Figure 12) was tested coupled to a magnetic micro-generator producing a 3-phase electrical output at 50 kHz. The magnitude of the generated line-to-line voltage varied from 1 to 3.3 V and the DC resistive load from 24 to 200 Ω. Under these typical operating conditions, the first electrical trials of the rectifier chip demonstrated its full functionality (including automatic start-up – Figure 13) and a conversion efficiency as high as 90 % (Figure 14).

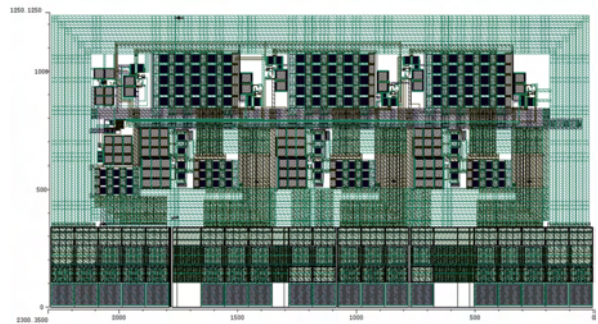

*Figure 11. Layout of the integrated µ-rectifier (2.9 mm²)*

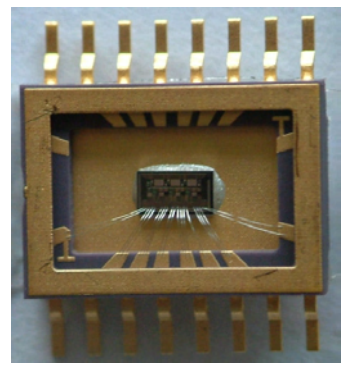

*Figure 12. Rectifier die in SOIC-16 testing package*

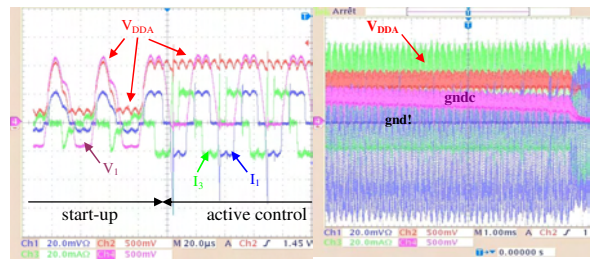

*Figure 13. Measured waveforms during the automatic start-up sequence*

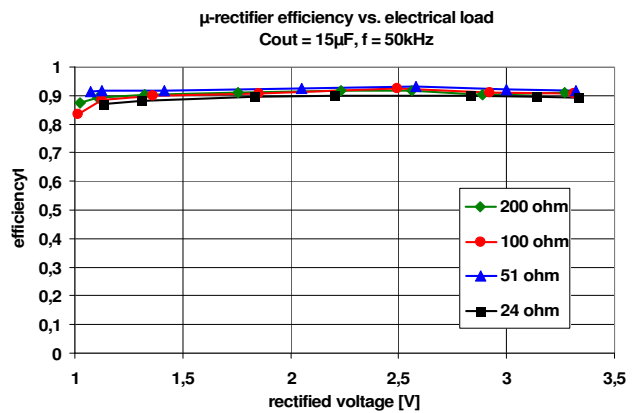

*Figure 14. Measured µ-rectifier efficiency (the µ-rectifier is supplied by the 3-phase magnetic µ-generator)*

This µ-rectifier has been cascaded with the boost converter presented above, to obtain a µ-converter which delivers a regulated output voltage of 5 V (Figure 15).

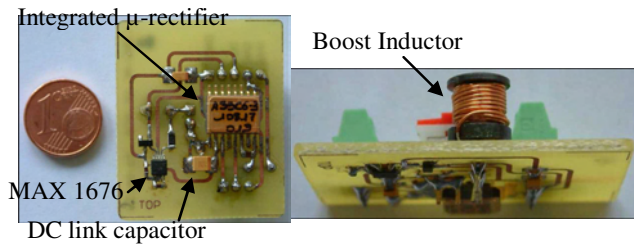

*Figure 15. Front and side views of the cascaded AC-DC µ-converter*

First tests indicate that the µ-converter is fully functional (Figure 16): it starts automatically, even for phase-to-phase input voltages as low as 1 V, and for large output loads. Power measurements show that the µ-converter efficiency is about 80 % (Figure 17), which is a good and demonstrating result for a low-voltage, low-power µ-converter.

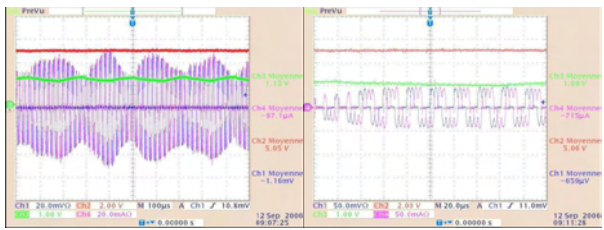

*Figure 16. Measurements on the 3-phase AC/DC µ-converter. $V_{out}$ = 5 V, $P_{out}$ = 25 mW.*

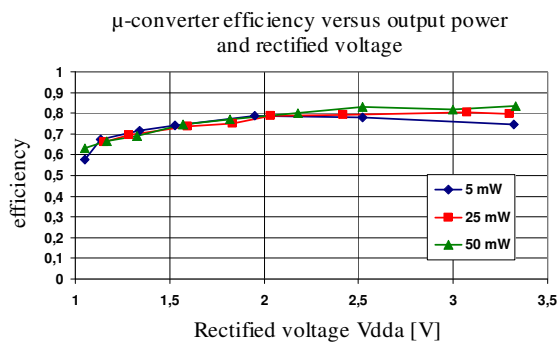

*Figure 17. Measured full AC-DC + DC/DC µ-converter efficiency versus $V_{DDA}$ for various loads.*

## 5. CONCLUSION

The paper has presented several power electronics topologies able to rectify and to regulate the electrical power provided by a low-voltage, low-power 3-phase micro-source. Based on design issues and performance levels but also for realistic first approach, a dual-stage structure has been selected. Specific design issues for stand alone operation, start-up sequence and/or under heavy capacitive loads have been underlined. Practical implementation with a real 3-phase micro-generator is presented in order to validate the work and the designs that have been carried out. The global efficiency of the two stages ranks around 70 to 82 %. Considering the application characteristics and a dual-stage topology these results are correct. The next steps are now directed towards the reduction of passive devices and the increase in efficiency, as we try to integrate all necessary functionalities and elements into a single-stage topology.

## 6. REFERENCES:

[1] A. S. Holmes, G. Hong, and K. R. Buffard, "Axial-flux permanent magnet machines for micropower generation," *J. Microelectromech. Syst.*, vol. 14, no. 1, pp. 54–62, 2005

[2] H. Jianyun, H. Yan, M. Hao, "High Efficient Rectifier Circuit Eliminating Threshold Voltage Drop for RFID Transponders", *in Proc: 6th International Conference On ASIC Proceedings*, Shangai, 2005, vol. 2, pp. 607-610

[3] H. Raisigel, O. Cugat and J. Delamare, "Permanent magnet planar micro-generators", *Sensors and Actuators A: Physical*, vol. 130–131, pp. 438–444, 2006.

[4] http://www.maxim-ic.com/

[5] www.linear.com

[6] M. Brunet, P. Dubreuil, E. Scheid, J-L. Sanchez, "Development of fabrication techniques for high-density integrated MIM capacitors in power conversion equipment", *SPIE MOEMS-MEMS Symposium Photonics West*, Jan 2006.

[7] S-C. Ó. Mathúna, T. O'Donnell, N. Wang, K. Rinne "Magnetics on Silicon: An Enabling Technology for Power Supply on Chip" *IEEE Transactions On Power Electronics*, VOL. 20, N° 3, May 2005

[8] V. Kursun, S.G. Narendra, V. K. De, E.G. Friedman, "Low-Voltage-Swing Monolithic DC-DC Conversion" *IEEE Transaction on Circuits and Systems—II: Express Briefs*, Vol. 51, N° 5, May 2004

[9] H. Raisigel, J.-C. Crebier, Y. Lembeye, J. Delamare, O. Cugat, "Autonomous, Low-Voltage, High Efficiency, CMOS Rectifier for Three-Phase Micro-Generators", *IEEE Transducers'07*, Lyon, France, June 10-14, 2007

[10] H. Raisigel, "Micro–générateur magnétique planaire et micro-convertisseur intégré", Ph.D. Thesis, INP Grenoble, France, Dec 2006

[11] http://asic.austriamicrosystems.com/

[12] http://cmp.imag.fr/